\documentstyle[aps,preprint]{revtex}
\tightenlines
\begin{document}
\title{\bf  BIPOLARON BINDING IN QUANTUM WIRES}
\author{E. P. Pokatilov}
\address{Laboratory Physics of Multilayer Structures, Department of
Theoretical Physics,\\ State University of Moldova, str. A. Mateevici, 60,
MD-2009 Kishinev, Republic of Moldova}
\author{V. M. Fomin$^a$, J. T. Devreese$^b$}
\address{Theoretische Fysica van de Vaste Stof, Departement
Natuurkunde,\\Universiteit Antwerpen (U.I.A.), Universiteitsplein 1,
B-2610 Antwerpen, Belgium}
\author{S. N. Balaban, and S. N. Klimin$^c$}
\address{Laboratory Physics of Multilayer Structures, Department of
Theoretical Physics,\\ State University of Moldova, str. A. Mateevici, 60,
MD-2009 Kishinev, Republic of Moldova}
\maketitle
\bigskip 
\centerline{Phys. Rev. B {\bf 61}, 2721-2728 (2000) 
\copyright 2000 The American Physical Society} 

\begin{abstract}
   A  theory  of  bipolaron  states in quantum wires with a  parabolic
potential   well   is	developed  applying  the  Feynman  variational
principle.  The  basic	parameters  of the bipolaron ground state (the
binding  energy,  the  number  of  phonons in the bipolaron cloud, the
effective mass, and the bipolaron radius) are studied as a function of
sizes of the potential well. Two cases are considered in detail:  a
cylindrical  quantum  wire and	a  planar  quantum wire. Analytical
expressions  for  the bipolaron parameters are obtained at large and small
sizes  of  the	quantum  well. It is shown that at $R\gg 1$
[where	$R$  means  the  radius  (halfwidth)  of  a cylindrical
(planar)  quantum wire, expressed in Feynman units], the
influence  of confinement on the bipolaron binding energy is described
by  the  function  $\sim 1/R^{2}$ for both cases, while at small sizes
this  influence  is  different	in each case.  In quantum wires, the
bipolaron binding energy $W\left( R\right) $ increases logarithmically
with	 decreasing   radius.	The   shapes  and  the	sizes  of a
nanostructure, which are favorable for observation of stable bipolaron
states, are determined.
\end{abstract}

\section{INTRODUCTION}

Landau's idea \cite{Landau} of the auto-localized state of a charge carrier
(polaron) in a homogeneous polar medium got a further development by Pekar
\cite{r1} who first studied a problem of a stable complex of two charge
carriers of the same sign (bipolaron). The bipolaron binding energy was
first calculated in Ref.~\onlinecite{Vinetskii}. The bipolaron problem was
widely discussed, see e.~g. Refs.~\onlinecite{r2,r3,r4,r5,r6,r7}. A detailed
outline of this subject is presented in the recent review \cite{r8}.

Dimensionless constants of the Coulomb interaction $U$ and of the
electron-phonon interaction $\alpha $ are related to each other by the
equation \cite{r8}:
\begin{equation}  \label{U-alpha}
U=\frac{\sqrt{2}\alpha }{1-\eta },
\end{equation}
where $\eta =\varepsilon _\infty /\varepsilon _0$ ($\varepsilon _0$ and $%
\varepsilon _\infty $ are static and optical dielectric constants,
respectively). Due to the fact that $\varepsilon _0>\varepsilon _\infty $,
the relation $U\geq \sqrt{2}\alpha $ follows. When the distance $l$ between
electrons is large or small compared with the characteristic polaron
radius $R_{{\rm p}}$ (see Ref.\,\onlinecite{r28}),
the phonon-mediated attraction between electrons
occurs to be weaker than the repulsion. At large distances $l\gg R_{{\rm p}}$%
, both interaction potentials have similar spatial dependences but the
Coulomb repulsion is stronger than the phonon-mediated attraction. In the
opposite case, $l\ll R_{{\rm p}}$, the Coulomb potential diverges at the
zero distance, while the phonon-mediated attraction is always finite.
Nevertheless, when two electrons move in such a way that the average distance
between them is of the same order as the polaron radius, the bipolaron can
be stable at $\alpha \gg 1$\ and $\eta \ll 1$. When two electrons are
together confined to a potential well, one can expect that the conditions of
the bipolaron stability may be improved for relevant sizes of the well.

Two new circumstances have stimulated the bipolaron theory: the progress in
the fabrication technology of mesoscopic nanostructures such as quasi-2D
(quantum wells and superlattices), quasi-1D (quantum wires), quasi-0D
(quantum dots), and the advancement of the hypothesis that bipolaron
excitations might play a role in processes occurring in the high-temperature
superconductors. The present research has been motivated also by the recent
advances in creation of nanocrystals with a strong ionic coupling \cite
{Hudgins}.

The basic bipolaron parameters are recalled in what follows. The bipolaron stability
region is determined by the inequality $W>0$ for the bipolaron binding
energy
\begin{equation}  \label{Wenergy}
W\equiv 2E_{{\rm p}}-E_{{\rm bip}}.
\end{equation}
Here $E_{{\rm p}}$ and $E_{{\rm bip}}$ are the free polaron and bipolaron
ground state energies, respectively.  From the equation
\begin{equation}  \label{Critical}
W\left( \alpha ,\eta ,{\cal R}\right) =0,
\end{equation}
where ${\cal R}$ denotes the set of parameters determining the shape and the
size of the confinement domain, the functions
$\alpha _c\left( \eta ,{\cal R }\right) $ and
$\eta _c\left( \alpha ,{\cal R}\right) $ describing the
boundaries of the bipolaron stability region are found, for fixed
$\eta$ and $\alpha$, respectively. According to different theoretical
treatments \cite{r2,r3,r4,r5,r6,r7,r8}	the
bipolaron binding energy is an increasing function of $\alpha $ and a
decreasing function of $\eta $. It will be shown that the function $\eta
_c\left( \alpha ,{\cal R}\right) $ starts from $\eta _c=0$ at $\alpha
=\alpha _{\min }\left( {\cal R}\right) \neq 0$, grows with increasing $%
\alpha $ and tends to the upper limit $\eta _{\max }$ at $\alpha \rightarrow
\infty $. The bipolaron stability region is then determined by the
inequalities $\alpha \geq \alpha _{\min }\left( {\cal R}\right) $ and $0\leq
\eta <\eta _c\left( \alpha ,{\cal R}\right) $.

Let us adduce typical values of the parameters $\alpha _{\min ,{\rm 3D}}$
and $\eta _{\max ,{\rm 3D}}$ of the bulk (3D) bipolaron: $\alpha _{\min ,
{\rm 3D}}=6.8$ and $\eta _{\max ,{\rm 3D}}=0.14$ were found by Verbist,
Peeters and Devreese \cite{Verb1,Verb2} and by Verbist, Smondyrev, Peeters
and Devreese~\cite{r9}. Adamowski~\cite{r5} obtained $\alpha _{\min ,{\rm 3D}
}=7.3$ and $\eta _{\max ,{\rm 3D}}=0.14$.
The bipolaron theory developed for pure 2D \cite{r10,r11} and 1D \cite{r11}
models shows that the bipolaron stability region broadens when the
dimensionality is reduced. For these systems, the following parameters were
obtained: $\alpha _{\min ,{\rm 2D}}=2.9$, $\eta _{\max ,{\rm 2D}}=0.158$
(Ref. \onlinecite{r10}); $\alpha _{\min ,{\rm 1D}}=0.9$,
$\eta _{\max ,{\rm 1D}}=0.764$ (Ref.~\onlinecite{r12}).
Bipolaron states were investigated in a quantum well~\cite{r13,r14} and in a
quantum wire~\cite{r15} as a function of the characteristic size of the
system. The polaron theory for a
quantum dot is developed in Refs.~\onlinecite{r16,r17,r18,pe}.

The goal of the present investigation is to determine the
bipolaron stability region and to study the basic parameters characterizing the
bipolaron ground state as a function of confinement. Two different types of
confinement are considered and compared to each other: (i)~a cylindrical
quantum wire of the radius $R$, where continuous transitions from 3D to 1D
are realized with decreasing $R$; (ii)~a planar quantum wire of the width
$L$, where a transition from 2D to 1D is realized with decreasing
$L$. A unique approach, namely the Feynman variational method
\cite{r26,r27}, is used
throughout the paper for both systems under analysis.

The paper is organized as follows. In Section~II, general formulae for
parameters of a bipolaron in quantum wires are
deduced. In Section~III, particular cases of cylindrical and planar quantum
wires are considered. The basic parameters of the bipolaron ground state are
obtained. Limiting cases of strong and weak confinement are studied in
detail. The obtained numerical and analytical results are discussed in
Section~IV. Section~V contains conclusions about the influence of
confinement on the bipolaron binding energy in quantum wires.

\section{General theory}

We analyze the bipolaron problem taking into account both the
electron-phonon interaction and the Coulomb repulsion between two electrons
confined to a quantum wire. The Lagrange function of the system is
\begin{eqnarray}
L &=&\sum_{i=1}^{{\rm D}}\sum_{n=1,2}\frac{m_{i}\dot{x}_{i,n}^{2}}{2}%
-\sum_{n=1,2}{\cal U}\left( {\bf r}_{n}\right) -\frac{e^{2}}{\varepsilon
_{\infty }\left| {\bf r}_{1}-{\bf r}_{2}\right| }  \nonumber \\
&&+\frac{1}{2}\sum_{{\bf k}}\left( \dot{w}_{{\bf k}}^{2}-\omega _{0}^{2}
w_{{\bf k}}^{2}\right) -\sum_{n=1,2}\sum_{{\bf k}}\gamma _{{\bf k}}
\left( {\bf r}_{n}\right) w_{{\bf k}},	\label{L}
\end{eqnarray}
where ${\bf r}_{n}\left( x_{1n},x_{2n},x_{3n}\right) $ is the radius vector
of the $n$-th electron ($n=1,2$); $m_{i}$ is the $ii$-th component ($i=1,2,3$)
of the diagonal band mass tensor, ${\cal U}\left( {\bf r}\right) $ is the
potential energy of an electron in the quantum wire, $w_{{\bf k}}$ are the
normal coordinates of longitudinal optical (LO) phonon modes. Here, the
parameter ${\rm D}$ determines the dimensionality of the electron
subsystem: ${\rm D}=3$ and $2$ for  cylindrical and planar quantum
wires, respectively.
Amplitudes of the electron-phonon interaction are taken in the Fr\"{o}hlich
form:
\begin{equation}
\gamma _{{\bf k}}\left( {\bf r}\right) =2\sqrt{\frac{2\pi \hbar \omega
_{0}\alpha }{V}}\frac{\omega _{0}}{k}\left( \frac{\hbar }{2\bar{m}\omega _{0}
}\right) ^{1/4}\exp \left( i{\bf kr}\right) ,  \label{gamma}
\end{equation}
where $\bar{m}\equiv \left( m_{1}m_{2}m_{3}\right) ^{1/3}$, $V$ is the
volume of the system, and the Fr\"{o}hlich constant
\begin{equation}
\alpha =\frac{e^{2}}{2\hbar \omega _{0}}\left( \frac{1}{\varepsilon _{\infty
}}-\frac{1}{\varepsilon _{0}}\right) \left( \frac{2\bar{m}\omega _{0}}{\hbar
}\right) ^{1/2}  \label{alpha}
\end{equation}
characterizes the strength of the coupling between an electron and bulk
polar LO phonons with the long-wavelength frequency $\omega _{0}$. In this
paper, the 3D phonon approximation is used, according to which the
interaction of an electron with both bulk-like and interface phonons is
replaced by that with 3D phonons. This often used approach is adequate
because any integral polaron or bipolaron effect, resulting from a summation
over all phonon modes, appears to be only weakly dependent on the details of
the phonon spectrum. It should be also mentioned that the system under
consideration simulates realistic structures with relatively smooth
interface barriers, where interface-like phonon modes can appear, which are
smoothly distributed in space rather than localized near a sharp
boundary, as is the case for interface modes.

In order to study the bipolaron problem at arbitrary values of $\alpha $,
the Feynman variational approach \cite{r27} is the most appropriate method.
The trial Lagrange function is written as
\begin{eqnarray}
L_{tr} &=&\frac{1}{2}\sum_{i=1}^{{\rm D}}\sum_{n=1,2}\left[ m_{i}\dot{x}%
_{i,n}^{2}+M_{i}\dot{X}_{i,n}^{2}-k_{i}\left( x_{i,n}-X_{i,n}\right)
^{2}-k_{i}^{\prime }\left( x_{i,n}-X_{i,\bar{n}}\right) ^{2}\right]
\nonumber \\
&&+\sum_{i=1}^{3}K_{i}\left( x_{i1}-x_{i2}\right) ^{2}-\sum_{n=1,2}{\cal W}%
\left( {\bf r}_{n}\right) ,  \label{L0}
\end{eqnarray}
where $X_{in}$ are coordinates of the $n$-th ``fictitious'' particle
($n=1,2$). This model
imitates the interaction of electrons with phonons and between each other by
elastic bonds as shown in Fig.~1. The masses $M_{i}$ and the force constants
$k_{i}$, $k_{i}^{\prime }$, $K_{i}$ play the role of variational parameters.
For $n=1$, $\bar{n}$ takes the value 2, and for $n=2$, $\bar{n}$ is equal to
1. The potential well ${\cal U}\left( {\bf r}\right) $ from Eq.\,(\ref{L}) is
simulated here by a parabolic function:
\begin{equation}
{\cal W}\left( {\bf r}\right) =\frac{1}{2}\sum_{i=1}^{q}m_{i}\Omega
_{i}^{2}x_{i}^{2}.  \label{parpoten}
\end{equation}
The index $q$ characterizes the dimensionality of confinement
and is determined as follows: for a planar quantum wire $q=1$
($\Omega _{1}\neq 0$ and $\Omega _{2}=0$) and
for a cylindrical quantum wire $q=2$ ($\Omega _{1}\neq 0$,
$\Omega_{2}\neq 0$, and $\Omega _{3}=0$).

The basis of the Feynman variational method is the Jensen-Feynman inequality
\cite{r27}:
\begin{equation}  \label{FeyJen}
\left\langle \exp \left( S-S_{tr}\right) \right\rangle _{S_{tr}}\ge \exp
\left\langle S-S_{tr}\right\rangle _{S_{tr}},
\end{equation}
where the angular brackets denote averaging over electron paths:
\begin{equation}
\left\langle G\right\rangle _{S_{tr}}=\frac{{\rm Tr}\int {\rm D}{\bf r}G%
\left[ {\bf r}\right] \exp \left( S_{tr}\right) }{{\rm Tr}\int {\rm D}{\bf r}%
\exp \left( S_{tr}\right) }.
\end{equation}
Here $S$ and $S_{tr}$ are the electron action functionals obtained after
integration over phonon variables and over coordinates of
``fictitious'' particles, respectively. At low temperatures, the variational bipolaron
energy is calculated using the expression
\begin{equation}  \label{E}
E_{bip}=E_{tr}-\lim _{\beta \rightarrow \infty }\frac{\left\langle
S-S_{tr}\right\rangle _{S_{tr}}}\beta ,
\end{equation}
where $E_{tr}$ is the ground state energy of the trial system with the
Lagrangian (\ref{L0}), $\beta =1/k_BT$ is the inverse temperature.

The trial Lagrange function (\ref{L0}) consists of ${\rm D}$ independent
parts: $L_{tr}=\sum\limits_{i=1}^{{\rm D}}L_i$. Each part $L_i$ is a
function of four variables $x_{i1}$, $x_{i2}$, $X_{i1}$, $X_{i2}$. Let us
introduce unified denotations for coordinates of electrons and of
``fictitious'' particles: $\tilde x_{i1}=x_{i1}$, $\tilde x_{i2}=x_{i2}$, $\tilde x%
_{i3}=X_{i1}$, $\tilde x_{i4}=X_{i2}$. It follows from the form of the trial
Lagrangian (\ref{L0}) with Eq.\,(\ref{parpoten}) that the groups of variables
$\tilde x_{ij}$ with different indices $i$ are dynamically independent from
each other. They are related to normal variables $\xi _{ij}$ by the unitary
transformation:
\begin{equation}  \label{transform}
\tilde x_{ij}=\sum_{j^{\prime }=1}^4d_{i,jj^{\prime }}\xi _{ij^{\prime
}},\quad i=1,\dots ,{\rm D}
\end{equation}
with $4\times 4$ matrices $\left\| d_{i,jj^{\prime }}\right\| $ ($%
j,j^{\prime }=1,\dots ,4$). From the equations of motion for the group of
coordinates $\tilde x_{ij}$ ($j=1,\dots ,4$) with a fixed $i$, the
following eigenfrequencies are obtained:
\begin{eqnarray}
\omega _{ij}^2&=&\frac 12\left\{ \left(1+\frac{M_i}{m_i}\right)v_i^2+\Omega_i^2-
(-1)^j\sqrt{\left[\left(1-\frac{M_i}{m_i}\right)v_i^2-\Omega _i^2\right]^2 +4
\frac{M_i}{m_i}v_i^4}\right\},\quad j=1,2,  \nonumber \\
\omega _{ij}^2&=&\frac 12\left\{ \left(1+\frac{M_i}{m_i}\right)v_i^2+\Omega_i^2-2
\frac{K_i}{m_i}- (-1)^j\sqrt{\left[\left(1-\frac{M_i}{m_i}\right)v_i^2-\Omega _i^2+ 2
\frac{K_i}{m_i}\right]^2 +4\frac{(k_i-k^{\prime}_i)^2}{m_iM_i}}\right\},  \nonumber
\\
&&\qquad\qquad\qquad\qquad\qquad\qquad\qquad\qquad\qquad\qquad
\qquad\qquad\qquad\qquad\qquad j=3,4,  \label{omega14}
\end{eqnarray}
where $v_i^2=(k_i+k_i^{\prime })/M_i$. Matrix elements of the unitary
transformation (\ref{transform}) are
\[
d_{i,11}^2=\frac{\omega _{i1}^2-v_i^2}{2\left( \omega _{i1}^2-\omega
_{i2}^2\right) },\quad d_{i,12}^2=\frac{v_i^2-\omega _{i2}^2}{2\left( \omega
_{i1}^2-\omega _{i2}^2\right) },\quad d_{i,13}^2=\frac{\omega _{i3}^2-v_i^2}{%
2\left( \omega _{i3}^2-\omega _{i4}^2\right) },\quad d_{i,14}^2=\frac{%
v_i^2-\omega _{i4}^2}{2\left( \omega _{i3}^2-\omega _{i4}^2\right) },
\]
\begin{equation}  \label{Elements}
d_{i,2j^{\prime }}=s_{j^{\prime }}d_{i,1j^{\prime }};\quad d_{i,3j^{\prime
}}=\frac{k_i+s_{j^{\prime }}k_i^{\prime }}{M_i\left( v_i^2-\omega
_{ij^{\prime }}^2\right) }d_{i,1j^{\prime }};\quad d_{i,4j^{\prime
}}=s_{j^{\prime }}d_{i,3j^{\prime }};
\end{equation}
\[
s_j=1\;(j=1,2);\;s_j=-1\;(j=3,4).
\]

Note that the elastic repulsion imitating the Coulomb interaction
gives a contribution to the eigenfrequencies with
$j=3$ and $4$ through the force constants $K_i$.
It is easy to see from Eq.\,(\ref{omega14}) that under the conditions
of a strong confinement along the $i$-th coordinate axis the
eigenfrequencies with $j=1$ and $3$
corresponding to the motion of the bipolaron
along this axis  as a whole are determined mainly by the
parameter $\Omega_i$.

The action functionals $S$ and $S_{tr}$ in Eqs.\,(\ref{FeyJen}) to (\ref{E})
contain the potential energies ${\cal U}$ and ${\cal W}$, respectively.
Though the shape of a real potential ${\cal U}$ may differ from that of the model
quadratic potential (\ref{parpoten}),  the averaged difference
$\left\langle {\cal U}-{\cal W}\right\rangle_{S_{tr}}$ can be omitted as far
as it is small when compared to the rest of
$\left\langle S-S_{tr}\right\rangle _{S_{tr}}/\beta$.

The averaging procedure in Eq.\,(\ref{E}) is carried out by the path
integration and leads to the following form of the variational bipolaron
energy:
\begin{equation}  \label{Ebip}
E_{{\rm bip}}=\sum_{i=1}^{{\rm D}}B_i+C+P.
\end{equation}
Here the terms $B_i$ include the averaged kinetic energies of two electrons and
of two ``fictitious'' particles as well as the averaged potential energy of
the elastic interaction of these four particles:
\begin{equation}  \label{Ti}
B_i=\frac 12\sum_{j=1}^4\omega _{ij}\left( 1-\frac{\omega _{ij}^2-\Omega _i^2%
}{\omega _{ij}^2}d_{i,1j}^2\right) -v_i,\quad i=1,\dots ,{\rm D}.
\end{equation}
In Eqs.\,(\ref{Ebip}), (\ref{Ti}) and further on, the Feynman
units\cite{r26} are used:
$\hbar \omega _0$ for energies; $\omega _0$ for frequencies; and
$\left( \hbar /\bar m\omega _0\right) ^{1/2}$ for lengths.

The averaged potential energy of the Coulomb electron repulsion is
\begin{equation}  \label{C}
C=\frac \alpha {\left( 1-\eta \right) \pi ^2}{\cal K}_2\left( 0\right) ,
\end{equation}
and the averaged energy of the electron-phonon interaction is
\begin{equation}  \label{P}
P=-\frac \alpha {\pi ^2}\sum_{n=1,2}\int\limits_0^\infty d\tau \,{\rm e}%
^{-\tau }{\cal K}_n\left( \tau \right) ,
\end{equation}
where
\begin{equation}  \label{K}
{\cal K}_n\left( \tau \right) =\int\limits_{-\infty }^\infty \frac 1{k^2}%
\exp \left[ -\sum\limits_{i=1}^{{\rm D}}k_i^2A_{in}\left( \tau \right) %
\right] \prod_{i=1}^{{\rm D}}dk_i.
\end{equation}
The functions $A_{in}\left( \tau \right) $ are determined as follows:
\begin{equation}  \label{Ain}
A_{in}\left( \tau \right) =\frac{\bar m}{m_i}\left[ \sum_{j=1,2}\frac{%
d_{i,1j}^2}{\omega _{ij}}\left( 1-{\rm e}^{-\omega _{ij}\tau }\right)
+\sum_{j=3,4}\frac{d_{i,1j}^2}{\omega _{ij}}\left( 1+\left( -1\right) ^n{\rm %
e}^{-\omega _{ij}\tau }\right) \right] ,\quad n=1,2.
\end{equation}
In order to find the bipolaron energy, it is necessary to minimize the
function $E_{{\rm bip}}$ given by Eq.\,(\ref{Ebip}) over twelve independent
variational parameters $\omega _{ij},$ $i=1\dots 3,$ $j=1\dots 4$, which are
used instead of the masses $M_i$ and the force constants $k_i$, $k_i^{\prime
}$, $K_i$.

From Eq.\,(\ref{omega14}) for the eigenfrequencies, the expression for
components of the diagonal tensor of relative bipolaron effective mass is
deduced straightforwardly:
\begin{equation}  \label{mbip}
\frac{\left( m_{{\rm bip}}\right) _i}{m_i}\equiv 2\left( \frac{ M_i}{m_i}%
+1\right) =\frac{2\left( \omega _{i1}^2+\omega _{i2}^2-\Omega _i^2\right) }{%
v_i^2},
\end{equation}
where the values of parameters $\omega _{i1}$, $\omega _{i2}$, $v_i$ are
taken which provide the bipolaron energy.

The number of phonons in the bipolaron cloud is determined by the general
expression of Ref.\onlinecite{r28}
\begin{equation}  \label{Nph}
N_{{\rm ph}}=\left\langle \frac{\partial S}{\partial \left( \hbar \omega
_0\right) }\right\rangle _{S_{tr}},
\end{equation}
which gives in the case under consideration:
\begin{equation}  \label{Nph1}
N_{{\rm ph}}=\frac \alpha {\pi ^2}\sum_{n=1,2}\int\limits_0^\infty {\cal K}%
_n\left( \tau \right) {\rm e}^{-\tau }\tau \,d\tau .
\end{equation}
Calculations of the average number of phonons according to  Eq.\,(\ref{Nph1})
are performed using the results of
minimization of the bipolaron energy.

\section{Bipolaron in cylindrical and planar quantum wires}

\subsection{Variational problem}

Here we write down the	variational bipolaron energies	in the
cylindrical and planar quantum wires (see Fig.~2).
Hereafter, the following
denotation for the confinement parameter is used: $\Omega _{i}\equiv \Omega
_{\perp }$, $i=1,q$. The electron mass is taken to be isotropic,
i.~e. $m_1=m_2=m_3=m$.
From Eqs.\,(\ref{Ebip}) to (\ref{P})  we obtain
the variational bipolaron energy
\begin{equation}
E_{{\rm bip}}=B_{\perp }+B_{\parallel }+C+P,  \label{Ebip1}
\end{equation}
where, in accordance with Eq.\,(\ref{Ti}),
\begin{equation}
B_{\parallel }=\frac{1}{2}\sum\limits_{j=1}^{3}\omega _{\parallel j}\left(
1-d_{{\rm D},1j}^{2}\right) -v_{\parallel },  \label{kinpar}
\end{equation}
\begin{equation}
B_{\perp }=\frac{q}{2}\left[ \sum\limits_{j=1}^{4}\omega _{\perp j}\left( 1-%
\frac{\omega _{\perp j}^{2}-\Omega _{\perp }^{2}}{\omega _{\perp j}^{2}}%
d_{1,1j}^{2}\right) -2v_{\perp }\right].   \label{kinetic}
\end{equation}
Here the frequencies of the motion along the $z$-axis (called below the
longitudinal motion)  are
\begin{equation}
\omega _{{\rm D}1}\equiv \omega _{\parallel 1}\;\omega _{{\rm D}%
2}=0,\;\omega _{{\rm D}j}\equiv \omega _{\parallel j},\;j=3,4,	\label{*}
\end{equation}
\[
v_{{\rm D}}\equiv v_{\parallel }
\]
and those of the motion in the $xy$-plane (the transverse motion) are
\begin{equation}
\omega _{ij}\equiv \omega _{\perp j},\;j=1,\dots ,4,\quad v_{i}\equiv
v_{\perp }
\end{equation}
with $i=1,2$ for $q=2$ and $i=1$ for $q=1$. In the case under
consideration, the integrations containing the function ${\cal K}_{n}\left(
\tau \right) $ are performed analytically. The calculation of the integrals
in Eqs.\,(\ref{C}) and (\ref{P}) yields the averaged potential energy of the
Coulomb repulsion between electrons
\begin{equation}
C=\frac{\sqrt{2}U}{\sqrt{\pi A_{\parallel 2}(0)}}F_{q}\left( 1-\frac{%
A_{\perp 2}(0)}{A_{\parallel 2}(0)}\right)   \label{e18}
\end{equation}
and the averaged energy of the electron-phonon interaction
\begin{equation}
P=-\frac{2\alpha }{\sqrt{\pi }}\sum_{n=1,2}\int\limits_{0}^{\infty }d\tau
e^{-\tau }\frac{1}{\sqrt{A_{\parallel n}(\tau )}}F_{q}\left( 1-\frac{%
A_{\perp n}(\tau )}{A_{\parallel n}(\tau )}\right) ,  \label{e19}
\end{equation}
where
\[
A_{\parallel n}\left( \tau \right) =\sum\limits_{j=3,4}\frac{d_{{\rm D}%
,1j}^{2}}{\omega _{\parallel j}}\left[ 1+(-1)^{n}{\rm e}^{-\omega
_{\parallel j}\tau }\right] +\frac{d_{{\rm D},11}^{2}}{\omega _{\parallel 1}}%
\left( 1-{\rm e}^{-\omega _{\parallel 1}\tau }\right) +d_{{\rm D}%
,12}^{2}\tau ;
\]
\[
A_{\perp n}\left( \tau \right) =\sum_{j=1,2}\frac{d_{1,1j}^{2}}{\omega
_{\perp j}}\left( 1-{\rm e}^{-\omega _{\perp j}\tau }\right)
+\sum\limits_{j=3,4}\frac{d_{1,1j}^{2}}{\omega _{\perp j}}\left[ 1+(-1)^{n}%
{\rm e}^{-\omega _{\perp j}\tau }\right] ,\;n=1,2;
\]
\begin{equation}
F_{q}(x)=\left\{
\begin{array}{cl}
\frac{\tanh ^{-1}\sqrt{x}}{\sqrt{x}},~ & q=2; \\
\int\limits_{0}^{\frac{\pi }{2}}\frac{d\varphi }{\left( 1-x\sin ^{2}\varphi
\right) ^{1/2}},~ & q=1.
\end{array}
\right.   \label{F(x)}
\end{equation}
The minimization of the variational bipolaron energy $E_{{\rm bip}}$
determined by Eqs.\,(\ref{Ebip1}) to (\ref{F(x)}) is carried out with respect
to eight variational parameters $v_{\parallel }$, $\omega _{\perp 2}$ ,
$\omega _{\parallel j},\;\omega _{\perp j}\;(j=1,3,4)$. Setting
$k_{i}^{\prime }$ and $K_{i}$ equal to zero in these formulas, the twice
value of the polaron energy \cite{pe} is obtained from
Eq.\,(\ref{Ebip1}). The binding energy is
then found according to Eq.\,(\ref{Wenergy}). Results of the calculation of $W
$ as a function of the quantum wire radius $R=\Omega _{\perp }^{-1/2}$ are
presented in Fig.~3 for different values of $\alpha $. Then the functions
$\alpha _{min}\left(R\right)\equiv\alpha _{c}\left(\eta=0,R\right)$ (Fig.~4)
and $\eta _{c}\left( R,\alpha \right) $ are obtained. The latter
function is used for calculation of the critical value of the Coulomb
repulsion constant $U_{c}\left( \alpha \right) $ in order to describe the
bipolaron stability region shown in Fig.~5. From Eq.\,(\ref{mbip}), taking
into account Eq.\,(\ref{*}), the relative bipolaron effective mass of the
longitudinal motion is derived as
\begin{equation}
\frac{\left( m_{{\rm bip}}\right) _{\parallel }}{m}=\frac{2\omega
_{\parallel 1}^{2}}{v_{\parallel }^{2}}.  \label{mbip1}
\end{equation}
Plots of the relative bipolaron mass as a function of $R$ are shown in
Fig.~6. A detailed discussion of the results will be given in Section~IV.

\subsection{Weak size quantization}

For a weak size quantization, the eigenfrequencies of the transverse motion
can be represented as expansion series in $\Omega _{\perp }$. In these
series, we take into account only two first terms:{\em \ }
\[
\omega _{\perp 1}^2=\omega _{\parallel 1}^2+\Omega _{\perp }^2\frac{\omega
_{\parallel 1}^2-v_{\parallel }^2}{\omega _{\parallel 1}^2}+O(\Omega _{\perp
}^4),\quad \omega _{\perp 2}^2=\Omega _{\perp }^2\frac{v_{\parallel }^2}{%
\omega _{\parallel 1}^2}+O(\Omega _{\perp }^4),
\]
\begin{equation}
\omega _{\perp 3}^2=\omega _{\parallel 3}^2+\Omega _{\perp }^2\frac{\omega
_{\parallel 3}^2-v_{\parallel }^2}{\omega _{\parallel 3}^2-\omega
_{\parallel 4}^2}+O(\Omega _{\perp }^4),\quad \omega _{\perp 4}^2=\omega
_{\parallel 4}^2+\Omega _{\perp }^2\frac{v_{\parallel }^2-\omega _{\parallel
4}^2}{\omega _{\parallel 3}^2-\omega _{\parallel 4}^2}+O(\Omega _{\perp }^4).
\end{equation}
As a consequence, Eq.\,(\ref{kinetic}) takes on the form
\begin{equation}  \label{Tperp}
B_{\perp }=q\left[B_{\parallel }+ \Omega _{\perp }\frac{v_{\parallel }}{%
4\omega _{\parallel 1}} \frac{3\omega _{\parallel 1}^2-v_{\parallel }^2} {%
\omega _{\parallel 1}^2}\right] +O(\Omega _{\perp}^2),
\end{equation}
and the parameters in the right-hand side of Eq.\,(\ref{e19}) satisfy the
relations
\begin{equation}  \label{LA}
A_{\perp n}(\tau)=A_{\parallel n}(\tau)-\frac 14\Omega _{\perp }\tau ^2\frac{%
v_{\parallel }^3}{\omega _{\parallel 1}^3}+O(\Omega _{\perp }^2),\quad n=1,2.
\end{equation}
Substituting the expression (\ref{LA}) in Eqs.\,(\ref{e18}), (\ref
{e19}), we obtain the energy of the Coulomb repulsion
\begin{equation}  \label{Coulomb}
C=\frac{\sqrt{2}U}{\sqrt{\pi }}\frac{f_{1q}}{\sqrt{A_{\parallel 12}(0)}}%
+O(\Omega _{\perp }^2),
\end{equation}
and the energy of the electron-phonon interaction
\begin{equation}  \label{e-ph}
P=-\frac{2\alpha }{\sqrt{\pi }}\sum_{n=1,2}\int\limits_0^\infty d\tau \,{\rm %
e}^{-\tau }\left[ \frac{f_{1q}}{\sqrt{A_{\parallel n}(\tau)}}+\frac{\Omega
_{\perp }}4\tau ^2\frac{v_{\parallel }^3}{\omega _{\parallel 1}^3}\frac{%
f_{2q}}{\sqrt{A_{\parallel n}^3(\tau)}}\right] +O(\Omega _{\perp }^2),
\end{equation}
where
\[
\quad f_{1q}=\left\{
\begin{array}{c}
1,\ q=2; \\
\frac \pi 2,\ q=1,
\end{array}
\right. \quad f_{2q}=\left\{
\begin{array}{c}
\frac 13,\ q=2; \\
\frac \pi 8,\ q=1.
\end{array}
\right.
\]

The bipolaron energy in this limiting case can be represented as
\[
E_{{\rm bip}}=E_{{\rm bip}}^0+\Delta E_{{\rm bip}}+O(\Omega _{\perp }^2),
\]
where $E_{{\rm bip}}^0$ is the bipolaron energy in three or two dimensions
for ${\rm D}=3$ or ${\rm D}=2$, respectively.
The confinement-induced shift
of the bipolaron energy  $\Delta E_{{\rm bip}}$ is given by
\begin{equation}  \label{e21}
\Delta E_{bip}=\Omega _{\perp }\left\{ q\frac{v_{\parallel }}{4\omega _3}%
\frac{3\omega _{\parallel 1}^2-v_{\parallel }^2}{\omega _{\parallel 1}^2}-%
\frac{2\alpha }{\sqrt{\pi }}f_{2q}\frac{v_{\parallel }^3}{4\omega
_{\parallel 1}^2}\int\limits_0^\infty d\tau e^{-\tau }\tau ^2\left( \frac 1{%
\sqrt{A_{\parallel 11}^3(\tau)}}+\frac 1{\sqrt{A_{\parallel 12}^3(\tau)}}%
\right) \right\} .
\end{equation}

It is worth mentioning, that in the strong coupling regime, the integrals in Eq.\,(\ref{e21}) are calculated
analytically, and the minimization of this variational function with respect
to the frequencies is performed explicitly. For this purpose we use
the results of Ref.~\onlinecite{Smondyrev}, where
the following analytical expressions for frequencies are obtained:
$\omega _{\parallel i}=\alpha ^2\widetilde{\omega }_i$ (for $i=1,3$),
$\omega _{\parallel 4}=1,v_{\parallel }=1$, where
\begin{equation}  \label{par1}
\widetilde{\omega }_1=\frac{128}{9\pi }\theta _{{\rm D}}\frac{\left[ 1-\zeta
^2(U)\right] ^4}{\zeta ^2(U)},\ \widetilde{\omega }_3=\frac{128}{9\pi }%
\theta _{{\rm D}}\left[ 1-\zeta ^2(U)\right] ^3,
\end{equation}
\[
\zeta (U)=\frac U{16\alpha }+\frac 12\sqrt{2+\left( \frac U{8\alpha }\right)
^2},\ \theta _{{\rm D}}=\left\{
\begin{array}{ll}
1,\qquad \  & {\rm D}=3; \\
\left( \frac{3\pi }4\right) ^2,\  & {\rm D}=2.
\end{array}
\right.
\]
Using these frequencies, we find
the confinement-induced shift of the bipolaron energy to be
\begin{equation}  \label{e22}
\Delta E_{{\rm bip}}=q\frac{\Omega _{\perp }} {2\alpha ^2\widetilde{\omega }%
_3}.
\end{equation}
This result differs qualitatively from that deduced in Ref.~\cite{BD96} for
a bipolaron in a weak magnetic field, where the cyclotron frequency $\omega
_c$ plays the role of $\Omega $. Namely, as distinct from Eq.\,(\ref{e22}),
in the equation from Ref.~\cite{BD96} for the first-order correction to the
bipolaron energy $\alpha ^2$ stands instead of $\alpha ^4$.

It is important to note, that this positive correction to the bipolaron
energy due to the confinement is less than the twice value of the respective
correction to the polaron energy \cite{pe}. The confinement-induced
variation of the bipolaron binding energy obeys the inequality
\begin{equation}  \label{dW}
\Delta W=\frac q2\frac \Omega\perp {\alpha ^2}\left[ \frac{9\pi } {2\theta_D}-%
\frac 1{\widetilde{\omega }_3}\right] >0.
\end{equation}
Thus, the enhancement of the bipolaron binding takes place due to the
confinement.

\subsection{Strong size quantization}

In the limiting case of a strong size quantization, the terms of the order
of $\Omega_\perp ^2$ play a determining role in Eq.\,(\ref{Ebip}). For the
frequencies of the transverse motion, the expansion in inverse powers of $%
\Omega ^2$ gives:

\begin{eqnarray}
\omega _{\perp 1}^2&=&\Omega_\perp^2+\frac{M_1}{m}v_\perp^2+O(\Omega_%
\perp^{-2}), \quad\quad\quad\quad\ \omega _{\perp
2}^2=v_\perp^2+O(\Omega_\perp^{-2}),  \nonumber \\
\omega _{\perp 3}^2 & = & \Omega_\perp^2+\frac{M_1}{m}v_\perp^2-2\frac{%
K_\perp}m+O(\Omega_\perp^{-2}), \quad \omega _{\perp 4}^2 =
v_\perp^2+O(\Omega_\perp^{-2}).  \label{e23}
\end{eqnarray}
Consequently, the bipolaron ground state energy is described by the
expression
\begin{eqnarray}
E_{{\rm bip}} & = & q\Omega_\perp +T_{\parallel }+ \frac {U}{\sqrt{2 \pi
A_{\parallel 2}(0)}}\ln \left( \frac{16}{q^2} \Omega_\perp A_{\parallel
2}(0)\right)  \nonumber \\
& & -\frac \alpha {\sqrt{\pi}}\sum\limits_{n=1,2} \int\limits_0^\infty d\tau
e^{-\tau } \frac{1}{\sqrt{A_{\parallel n}(\tau)}} \ln \left( \frac{16}{q^2}%
\Omega_\perp A_{\parallel n}(\tau)\right).  \label{e24}
\end{eqnarray}
The first term in the right-hand side of Eq.\,(\ref{e24}) is the energy of
two electrons in the parabolic potential.  The last three terms in the
right-hand side of Eq.\,(\ref{e24}) are due to the electron-phonon
and Coulomb interactions.
In the strong coupling regime, $\omega _{\parallel i}=\alpha ^2\tilde \omega
_i$ ($i=1,3$)$;$ $\omega _{\parallel 4}$ and $v_{\parallel }$ are
proportional to $\alpha ^0$ with coefficients which are functions of $\Omega
_{\perp }$. Then omitting the terms of the order of
$\alpha ^0$ in the last three terms of the variational bipolaron energy
(\ref{e24}) we obtain
\begin{eqnarray}
E_{{\rm bip}} & = & q\Omega_\perp +\alpha^2 \left[ \frac{\widetilde{\omega }%
_1+\widetilde{\omega }_3}{4}+ \frac{U \sqrt{\widetilde{\omega }_1}} {\alpha%
\sqrt{2\pi }}\ln \left( \frac{16}{q^2} \frac {\Omega_\perp} {\alpha^2%
\widetilde{\omega }_1}\right) \right.  \nonumber \\
& & \left. -\frac{2\sqrt{2}}{\sqrt{\pi }} \sqrt{\frac{\widetilde{\omega }_1%
\widetilde{\omega }_3} {\widetilde{\omega }_1+\widetilde{\omega }_3}}
\ln\left( \frac{8}{q^2}\frac {\Omega_\perp} {\alpha ^2}\left( \frac 1{%
\widetilde{\omega }_1}+ \frac 1{\widetilde{\omega }_3}\right) \right) \right]%
.  \label{e25}
\end{eqnarray}
Note, that the second term in the right-hand side of
Eq.\,(\ref{e25}) is proportional to $\alpha ^2$, as is expected in
the strong coupling regime. When replacing $\Omega _{\perp }\rightarrow
\omega _c$ at $q=2$, this polaronic term coincides with that of the
bipolaron variational energy in a strong magnetic field from Ref.~%
\onlinecite{BD96}. The dependence of the bipolaron binding energy on the
cylindrical confinement provides a possibility for
a controllable enhancement of the bipolaron binding by decreasing the
radius of a quantum wire.

Setting $U=0$ and $\tilde \omega =\tilde \omega _1=\tilde \omega _3$ in
Eq.\,(\ref{e25}), one obtains the twice variational polaron energy $2E_p$ with the
variational parameter $\tilde \omega $ (see Ref.~\onlinecite{pe}). The
polaron energy $E_p$ depends on the confinement parameter similarly
to $E_{{\rm bip}}$.
The binding energy $W$ (which is not written explicitly to save space)
increases logarithmically with increasing $\Omega _{\perp }$.

\section{DISCUSSION\ OF\ NUMERICAL\ RESULTS}

Beyond the framework of the limiting cases which allow an analytical
treatment as discussed above, the bipolaron stability is
studied using the following computational procedure. First, we evaluate the
bipolaron energy $E_{{\rm bip}}$ and the model bipolaron effective mass
defined as $m_{{\rm bip}}=2(M_{\parallel }+m)$. Second, the functions
$\alpha _{\min}\left( R\right) $ and $\eta _{c}\left( \alpha,R \right) $ are found
from Eq.\,(\ref{Critical}). The region of the Fr\"{o}hlich coupling constant
ranging from 2 to 4 is chosen for the numerical work in order to include the
values of $\alpha $ corresponding to some specific substances with small $%
\eta $ (for example, ${\rm TiO}_{2}$: $\alpha =2.03$, $\eta =0.035$ \cite
{Firsov}; TlCl: $\alpha =2.56$, $\eta =0.133$ \cite{Dev}; BaO: $\alpha =3.23$%
, $\eta =0.118$ \cite{Firsov}; LiBr: $\alpha =4.15$, $\eta =0.24$ \cite{Dev}%
).

Fig.~3 illustrates the size dependence of the bipolaron binding energy in
cylindrical and planar quantum wires, for $\alpha $ values corresponding to
the above-mentioned substances and for $\eta =0$. As seen from Fig.~3, the
bipolaron binding energy monotonously rises with increasing transverse
confinement [cf. Eqs.\,(\ref{Ebip1}) to (\ref{e19})].

In Fig.~4, the minimal value $\alpha _{\min}$ is represented as a function of $R$
and $L$ for cylindrical and planar quantum wires, correspondingly. In the
region of large $R$ and $L$, the minimal values $\alpha _{\min}$ for cylindrical
and planar quantum wires tend to the three-dimensional and two-dimensional
limits $\alpha _{\min,3{\rm D}}$ and $\alpha _{\min,2{\rm D}}$, respectively. When
$R$ and $L$ decrease from 1.0 to 0.1, a rapid diminution of
$\alpha_{\min}$ is
seen. Note that at small values of $R,L$ (which are, however, still
compatible with the continuum description), the bipolaron stability region
extends to small values of $\alpha $. Note that the bipolaron parameters for quantum
wires obtained in the formal limiting cases $R,L\rightarrow 0$ differ
substantially from those derived for the purely one-dimensional model \cite
{r12} (with 1D-electrons and 1D-phonons), which gives
$\alpha_{\min,1{\rm D}}=0.9$.

In Fig.~5, the ratio of the critical value of the Coulomb repulsion constant
to the Fr\"ohlich coupling constant $\alpha $,
\begin{equation}  \label{Ucrit}
\frac{U_c\left( \alpha \right) }\alpha =\frac{\sqrt{2}}{1-\eta _c\left(
\alpha \right) }
\end{equation}
is plotted as a function of $\alpha $ for various radii (ranging
from 0.01 to 20.0) of the cylindrical quantum wire. Since the parameter $%
\eta _c$ is non-negative, $U_c\left( \alpha \right) /\alpha $ cannot be less
than the value $\sqrt{2}$ (shown by the line $A$). When increasing $\alpha $%
, the right-hand side of Eq.\,(\ref{Ucrit}) tends to the three-dimensional
limit $\sqrt{2}/\left( 1-\eta _{c,3{\rm D}}\right) $, marked by the line $B$.
The physical sense of this trend consists in the following: when
increasing the electron-phonon coupling, the electron confinement to the
{\it parabolic potential} (\ref{parpoten}) is gradually replaced by the
confinement to the {\it polaronic potential} well. The regions of bipolaron
stability can exist only between the lines $A$ and $B$. The domain between a
curve $U_c\left( \alpha \right) /\alpha $ and the line $A$ is the bipolaron
stability region for a specific radius of the cylindrical quantum wire.
This figure illustrates clearly an enlargement of the bipolaron
stability region with decreasing the radius of the quantum wire. An
analogous dependence of the bipolaron stability region on the width takes
place for the planar quantum wire.

Bipolaron effective masses are represented in Fig.~6 as a function of the
dimensionless sizes $R$ and $L$ of the cylindrical and planar quantum wires.
The size dependence of the bipolaron effective mass is qualitatively similar
to that of the ground state energy but appears to be substantially more
pronounced. At small radii ($0.1\leq R\leq 0.2$), the bipolaron mass
strongly increases with decreasing $R$.

\section{CONCLUSIONS}

The conclusion of our analysis is that the confinement leads to an {\it %
enlargement of the bipolaron stability region} in cylindrical and planar
quantum wires as compared to the corresponding regions of infinite
three-dimensional and two-dimensional systems, respectively. For $R\sim 1$
or $L\sim 1$, the critical values $\alpha _c$ required for bipolaron
stability are close to those for TiO$_2$, TlCl, BaO and LiBr. For example,
according to Fig.~4b, the bipolaron stability region sets in at the width $L$
of the planar quantum wire of about 8 nm for parameters of TlCl. In this
view, manifestations of the bipolaron phenomena might be observed in the
technically achievable planar quantum wire structures.

The performed analytical and numerical analysis of the influence of
confinement on the bipolaron binding energy has shown that stable bipolaron
states are possible even for intermediate values of $\alpha $ ($\alpha \sim
2 $) and for not too small values of $\eta $ ($\eta \sim 0.1$) in
nanostructures whose sizes are of the same order as the polaron radius $R_{%
{\rm p}}$. Among the considered systems, the most favorable conditions for
the bipolaron stability take place in planar quantum wires, where the
binding energy $W$ increases monotonously (logarithmically) with
strengthening confinement. Nanostructures, whose sizes satisfy the
conditions of the bipolaron stability, seem to be achievable for the modern
technology.

\acknowledgements
We thank V. N. Gladilin for valuable discussions. This work has been
supported by the Interuniversitaire Attractiepolen --- Belgische Staat,
Diensten van de Eerste Minister --- Wetenschappelijke, technische en
culturele Aangelegenheden; Bijzonder Onderzoeksfonds (BOF) NOI of the
Universiteit Antwerpen; PHANTOMS Research Network; F.W.O.-V. projects Nos.
G.0287.95 and the W.O.G. WO.025.99N (Belgium).
S.N.K. acknowledges a financial support
from the UIA. E.P.P, S.N.K. and S.N.B. acknowledge with gratitude the kind
hospitality during their visits to the UIA in the framework of the common
research project supported by PHANTOMS.

\newpage

\begin{center}
{FIGURE CAPTIONS}
\end{center}

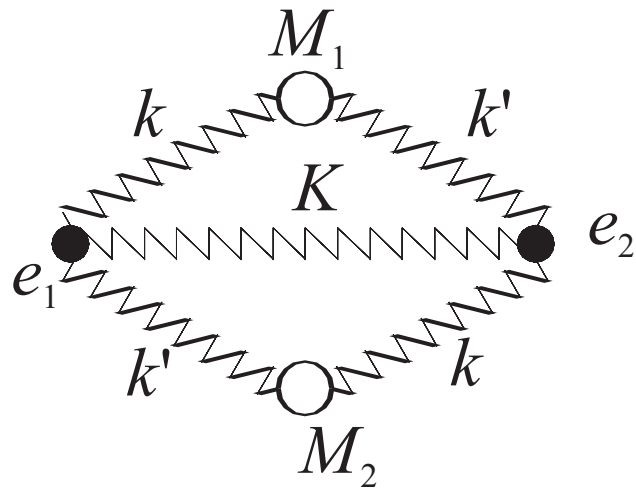
\noindent Fig.~1. A scheme of the trial system which contains two electrons
connected with two ``fictitious'' particles through the elastic attraction and
models the Coulomb interaction by the elastic repulsion.

\bigskip

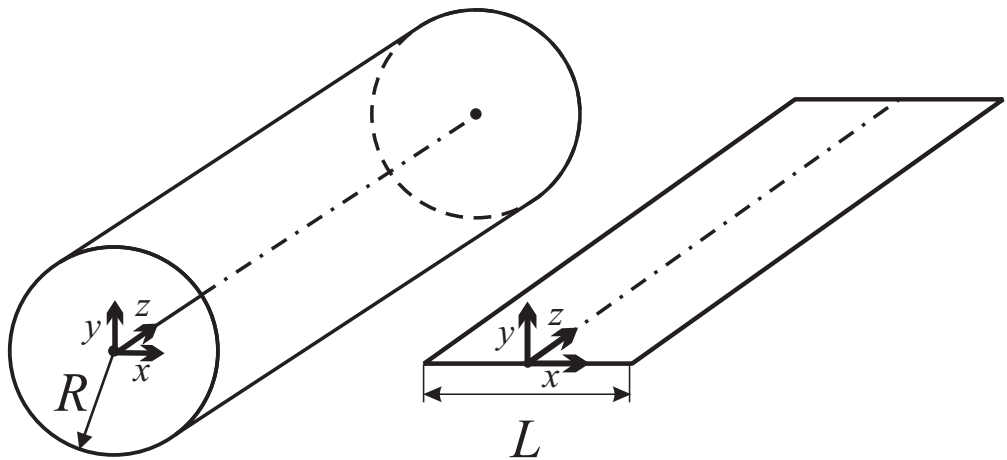
\noindent Fig.~2. A scheme of cylindrical (a) and planar (b) quantum wires.

\bigskip

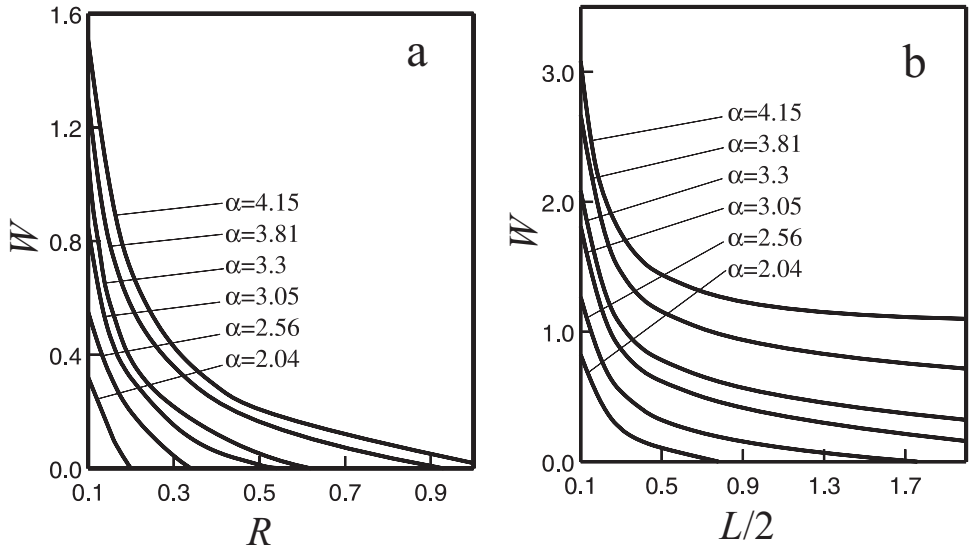
\noindent Fig.~3. The bipolaron binding energy $W$ in cylindrical (a) and
planar (b) quantum wires plotted versus the dimensionless radius $R$ and
width $L$, respectively.

\bigskip

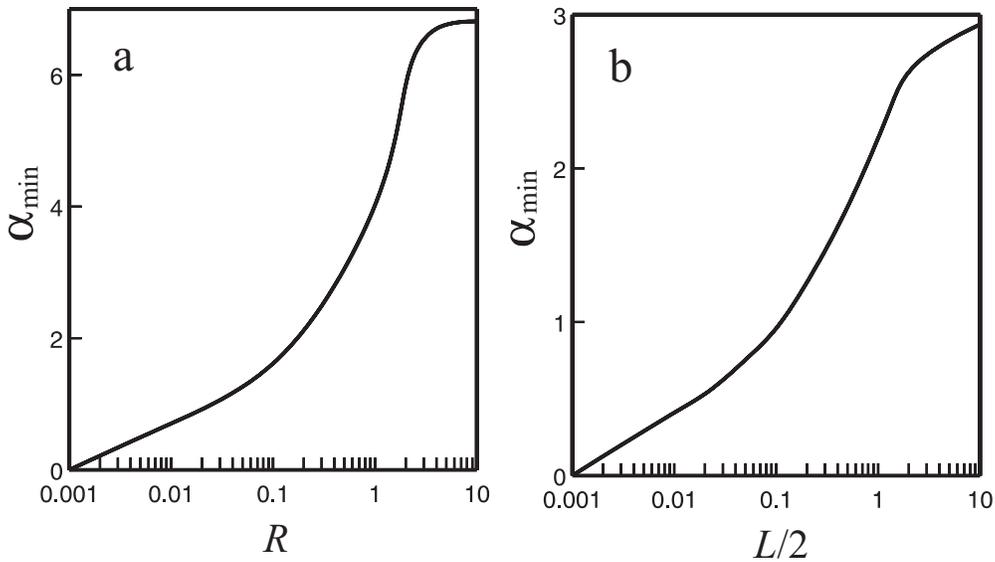
\noindent Fig.~4. The minimal value (at $\eta=0$) of the critical electron-phonon coupling
constant $\alpha _c$ plotted versus $R$ and $L$ in cylindrical (a) and
planar (b) quantum wires, respectively.

\bigskip

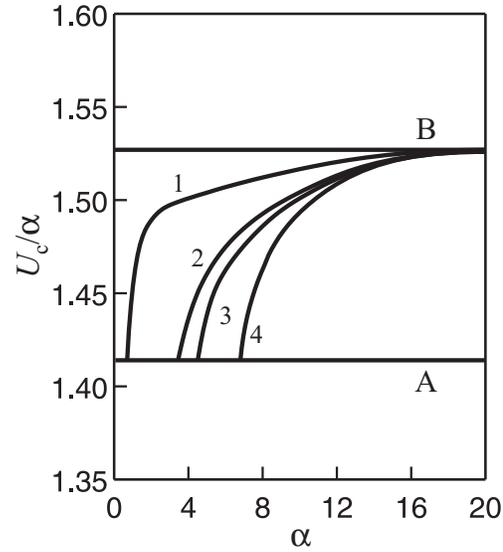
\noindent Fig.~5. The ratio of the critical Coulomb repulsion constant $U_c$
and $\alpha $, as a function of $\alpha $, in cylindrical quantum wires for $%
R=0.01$ (1), 0.5 (2), 1.0 (3), and 20.0 (4).

\bigskip

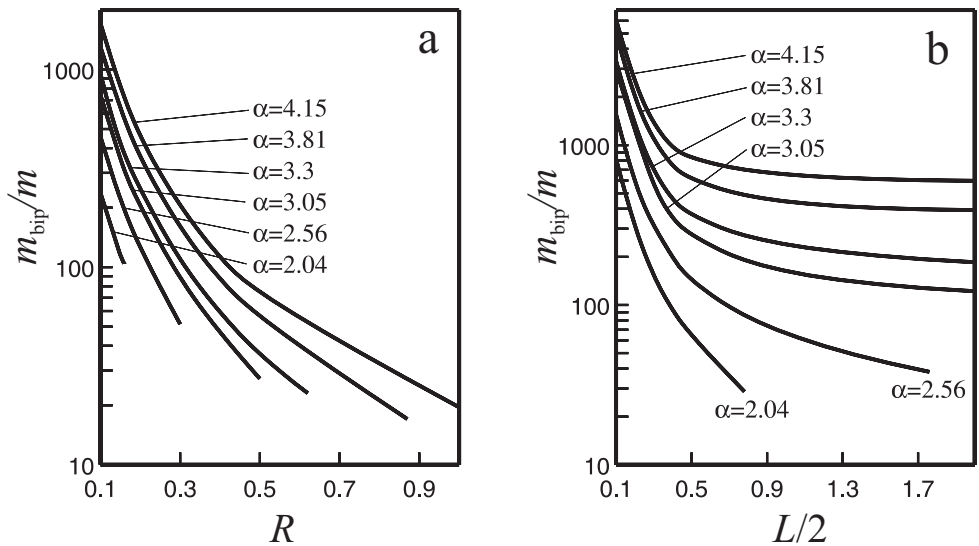
\noindent Fig.~6. The bipolaron effective mass
${(m_{\rm bip})}_\parallel/m$ in
cylindrical (a) and planar (b) quantum wires plotted versus $R$ and $L$,
respectively.  The curves for the effective mass are broken
off as the bipolaron becomes unstable.

\end{document}